\documentclass[lettersize,journal]{IEEEtran}
\usepackage{amsmath,amsfonts}
\usepackage{algorithmic}
\usepackage{algorithm}
\usepackage{array}
\usepackage[caption=false,font=normalsize,labelfont=sf,textfont=sf]{subfig}
\usepackage{textcomp}
\usepackage{stfloats}
\usepackage{url}
\usepackage{verbatim}
\usepackage{graphicx}
\usepackage{cite}
\usepackage{caption}
\usepackage{subcaption}
\usepackage{xcolor, color, soul}
\begin{document}

\title{Error Performance Analysis of UAV-Mounted RIS for NOMA Systems with Practical Constraints}

\author{Faical Khennoufa, Khelil Abdellatif, Ferdi Kara,~\IEEEmembership{Senior Member,~IEEE}, Halim Yanikomeroglu,~\IEEEmembership{Fellow,~IEEE}, Khaled Rabie,~\IEEEmembership{Senior Member,~IEEE}, Taissir Y. Elganimi,~\IEEEmembership{Senior Member,~IEEE}, Safia Beddiaf
\thanks{ F. Khennoufa is with LGEERE Laboratory, Department of Electrical Engineering, Echahid Hamma Lakhdar University, El-Oued, Algeria, and Non-Terrestrial Network (NTN) Laboratory, Department of Systems and Computer Engineering, Carleton University, Ottawa, K1S 5B6, ON, Canada, email: khennoufa-faical@univ-eloued.dz and FaicelKhennoufa@cunet.carleton.ca. A. Khelil and S. Beddiaf are with LGEERE Laboratory, Department of Electrical Engineering, Echahid Hamma Lakhdar University, El-Oued, Algeria, email: \{abdellatif-khelil, beddiaf-safia\}@univ-eloued.dz. F. Kara is with the Wireless Communication Technologies Laboratory (WCTLab), Department of Computer Engineering, Zonguldak Bulent Ecevit University, Zonguldak, Turkey, 67100,  e-mail: f.kara@beun.edu.tr. H. Yanikomeroglu is with Non-Terrestrial Network (NTN) Laboratory, Department of Systems and Computer Engineering, Carleton University, Ottawa, K1S 5B6, ON, Canada, email: halim@sce.carleton.ca. K. Rabie is with the Department of Engineering, Manchester Metropolitan University, Manchester M1 5GD, U.K., email: k.rabie@mmu.ac.uk. T. Y. Elganimi is with the Department of Electrical and Electronic Engineering, University of Tripoli, Libya, email: t.elganimi@uot.edu.ly.

{\color{black}This study is supported in part by the Study in Canada Scholarship (SICS) by Global Affairs Canada, and in part by the Discovery Grant RGPIN-2022-05231 from the Natural Sciences and Engineering Research Council of Canada (NSERC).}
}
}

\markboth{Preparing paper for IEEE Communications Letters Journal 
}%
{Shell \MakeLowercase{\textit{et al.}}: A Sample Article Using IEEEtran.cls for IEEE Journals}

\maketitle
\begin{abstract}
Uncrewed aerial vehicles (UAVs) have attracted recent attention for sixth-generation (6G) networks due to their low cost and flexible deployment. In order to maximize the ever-increasing data rates, spectral efficiency, and wider coverage, technologies such as reconfigurable intelligent surface (RIS) and non-orthogonal multiple access (NOMA) are adapted with UAVs (UAV-RIS NOMA). However, the error performance of  UAV-RIS NOMA has not been considered, yet. In this letter, we investigate the error probability of UAV-RIS NOMA systems. We also consider the practical constraints of hardware impairments (HWI) at the transceivers, inter-cell interference (ICI), and imperfect successive interference cancellation (SIC). The analytical derivations are validated by Monte-Carlo simulations. Our results demonstrate that our proposed system achieves higher performance gain (more than 5 dB with increasing the number of RIS elements) with less error probability compared to UAVs without RIS. Moreover, it is found that the HWI, ICI, and imperfect SIC have shown a negative impact on the system performance. 
\end{abstract}


\section{Introduction}
\IEEEPARstart{D}{ue} to the unprecedented growth of punctual mobile data traffic and strict quality-of-service (QoS) requirements, the upcoming next generation of wireless networks are set to provide ultra-reliable and low-latency communication and immersive reality, etc, which will enable a variety of applications, such as multiple access techniques, reconfigurable intelligent surfaces (RISs), low-orbit satellites, and uncrewed aerial vehicle (UAV) networks \cite{alfattani2021aerial}.
In this regard, non-orthogonal multiple access (NOMA) offers high connection and low latency. In NOMA, the superposition-coding signal with different power levels is used for user signals \cite{bariah2023performance}. Besides, UAV communication networks are seen as an exciting option to increase the QoS in crowded areas and extend the coverage of current networks in difficult environments due to their low cost, line-of-sight (LoS) air-to-ground channel, and flexible deployment \cite{bariah2023performance}.
However, the flying duration of UAVs is strictly constrained preventing them from meeting the demands of long missions. To resolve this problem, researchers proposed several solutions, such as battery capacity increase, wireless charging, and solar energy. Recently, researchers have advocated the use of RISs in wireless networks to construct smart radio environments for energy-efficient and cost-effective implementations. The structure of RIS is made of a thin and flexible metasurface material that contains passive circuitry. It is able to intelligently control the propagation of the incident electromagnetic waves and reflect or refract these waves in a desired way, thereby turning the wireless channel into a controllable environment, and enhancing the overall system performance \cite{alfattani2021aerial,bariah2023performance}.

In the existing literature, several papers consider the scenario of RIS with UAV, but the RIS is fixed on the ground (e,g., wall of a tall building) in their assumptions as in 
\cite{agrawal2021performance}. In this regard, a new vision is proposed in \cite{alfattani2021aerial} in which UAVs can be adapted with RIS (referred to as UAV-RIS) and used for wireless communications as a reflective layer between ground base stations and ground devices. This trend has attracted a great deal of research attention in recent studies \cite{bariah2023performance,solanki2022performance} where the outage probability and ergodic spectral efficiency are investigated for UAV-RIS to serve NOMA ground devices. The optimal resource allocations of UAV-RIS with zero-forcing beamforming have been developed in \cite{li2021aerial}, where each UAV serves a number of ground device in its own cluster. In another work \cite{nguyen2022uav}, the maximization problem of UAV-RIS has been formulated by determining the best power control factors at the ground base station (consisting of a massive antenna) and the phase shifts of several RISs. On the other hand, the radio-frequency equipment is a critical component of the massive network of interrelated devices that enables communication between nodes. In practical communication systems, the radio-frequency component experiences a variety of hardware impairments (HWI) that compromise the overall system performance \cite{canbilen2022performance,beddiaf2022unified}. In addition, another issue in cellular networks affecting the system performance is inter-cell interference (ICI), where cell-edge devices or nodes are more exposed to signals from neighboring cells \cite{zhou2023outage}.

As we have mentioned above, UAV-RIS has been analyzed for the outage probability, ergodic spectral efficiency, and throughput in \cite{bariah2023performance,solanki2022performance,li2021aerial,nguyen2022uav}. To the best of the authors knowledge, the bit error rate (BER) of UAV-RIS with/without NOMA has not been investigated. In addition, it is important to take into account practical scenarios for the performance analysis, such as the effect of HWI, ICI, and imperfect successive interference cancellation (SIC). 
Based on the above discussions, in this letter, we investigate UAV-RIS for NOMA with a joint impact of HWI, ICI, and imperfect SIC, as being in practical implementations. Thus, the contributions of this letter are provided as follow:

\begin{itemize}
\item We consider the UAV-RIS-aided NOMA network scheme while assuming that the direct link is unavailable due to obstacles, the NOMA ground devices receive signals using UAV-RIS. We consider that the ideal and non-ideal HWI exist at all nodes in our system. In addition, we consider the ground devices suffer from ICI of the other cells (e.g., the signals forwarded from the UAV-RIS of the neighboring cells), which is the first investigation with HWI and ICI jointly in the literature for RIS. The presence of HWI and ICI affects the SIC process to detect the signals of NOMA users. Thus, the imperfect SIC is considered in our investigation.
\item We derive the BER of UAV-RIS under HWI, ICI, and imperfect SIC. Also, we obtain the upper bound of our proposed system. For the sake of fair comparison, we compare our results with two schemes: UAV without RIS and traditional UAV-RIS OMA and we validate our analysis with Monte-Carlo simulations.
\item The results show that the proposed scheme achieves higher performance gain compared to UAVs without RIS. Moreover, it is found that the HWI and ICI have a negative impact on the system performance, which are affects the SIC to detect the symbols at the receivers. 

\end{itemize}

\section{System and Channel models}
In this letter, we consider that the source (S) communicates with two NOMA users named U$_1$ and U$_2$ with the aid of UAV-RIS as illustrated in Fig. 1. In this scenario, we consider the following assumptions: 1) The users are equipped with a single antenna. In contrast, the UAV is equipped with RIS that has $N$ as a symbol reflecting elements. 2) The HWI exists at S, NOMA users, and the nodes of neighboring cells. 3) The direct communication between S and users is blocked due to large obstacles, and the NOMA users receive the S signal through the reflected link only. 4) The users experience interference from neighboring cells (i.e., UAV-RIS transmissions from other cells), called ICI.
{\color{black}Also, in the paper, the phase-shift matrix of the RIS is assumed to be discrete, with individual elements signifying phase shifts introduced by reflecting elements. This allows phase shifts to be independently adjusted, giving the RIS more control over the propagation of electromagnetic waves for improved wireless channel efficiency \cite{canbilen2022performance}.} Usually, the communication link involving S-UAV and UAV-U$_i$ considers the influence of both small-scale and large-scale fading. Large-scale fading is contingent on the distance existing between the transmitter and receiver \cite{singh2022noma}. In the downlink scenario, we take into account three-dimensional Cartesian coordinates $(x,y,z)$ for positioning UAV and NOMA users as presented in Fig. 1. We begin by identifying the location of the S, U$_1$, and U$_2$ as follows. The position coordinates of the S is $\mathsf{S}(-x_s,0,0)$, whereas the position coordinates of ground users U$_1$ and U$_2$ are indicated as $\mathsf{U_1} 
(x_1,0,-z_1)$ and $\mathsf{U_2} (x_2,0,z_2)$, respectively. The central position of the UAV's circular trajectory, characterized by a radius $r$ and altitude $T$, can be pinpointed to the coordinates $O'(0, h, 0)$. Supposing that $\varphi$ is the angle between the position of the UAV on the circle and the x-axis, we can easily indicate the position coordinates of the UAV as $\mathsf{UAV}(r.\cos(\varphi),r.\sin(\varphi),T)$. Based on this analysis, we calculate the Euclidean distance between S-UAV, UAV-U$_1$, and UAV-U$_2$, respectively, as $d_s=\sqrt{r^2+T^2+x_{s}^2+2rx_{s}\cos(\varphi)}$, $d_1=\sqrt{(r\cos(\varphi)-x_{1})^2+(r\sin(\varphi)+z_{1})^2+T^2}$, and $d_2=\sqrt{(r\cos(\varphi)-x_{2})^2+(r\sin(\varphi)-z_{2})^2+T^2}$. Additionally, we assume that small-scale fading between S-UAV and UAV-U$_i$ follows the Rayleigh distribution\footnote{We assume that there is no direct link between the BS-UAV and the UAV-U$_i$ because of the presence of tall buildings or other obstructions (i.e., possible obstruction of signals by natural features), which cause reflections, diffractions, and scattering of the signals. {\color{black}This paper uses Rayleigh fading to model the wireless communication environment for UAVs, providing a statistical representation of signal fluctuations caused by multipath propagation.}}.


\begin{figure}
\centering
\includegraphics[width=.8\columnwidth]{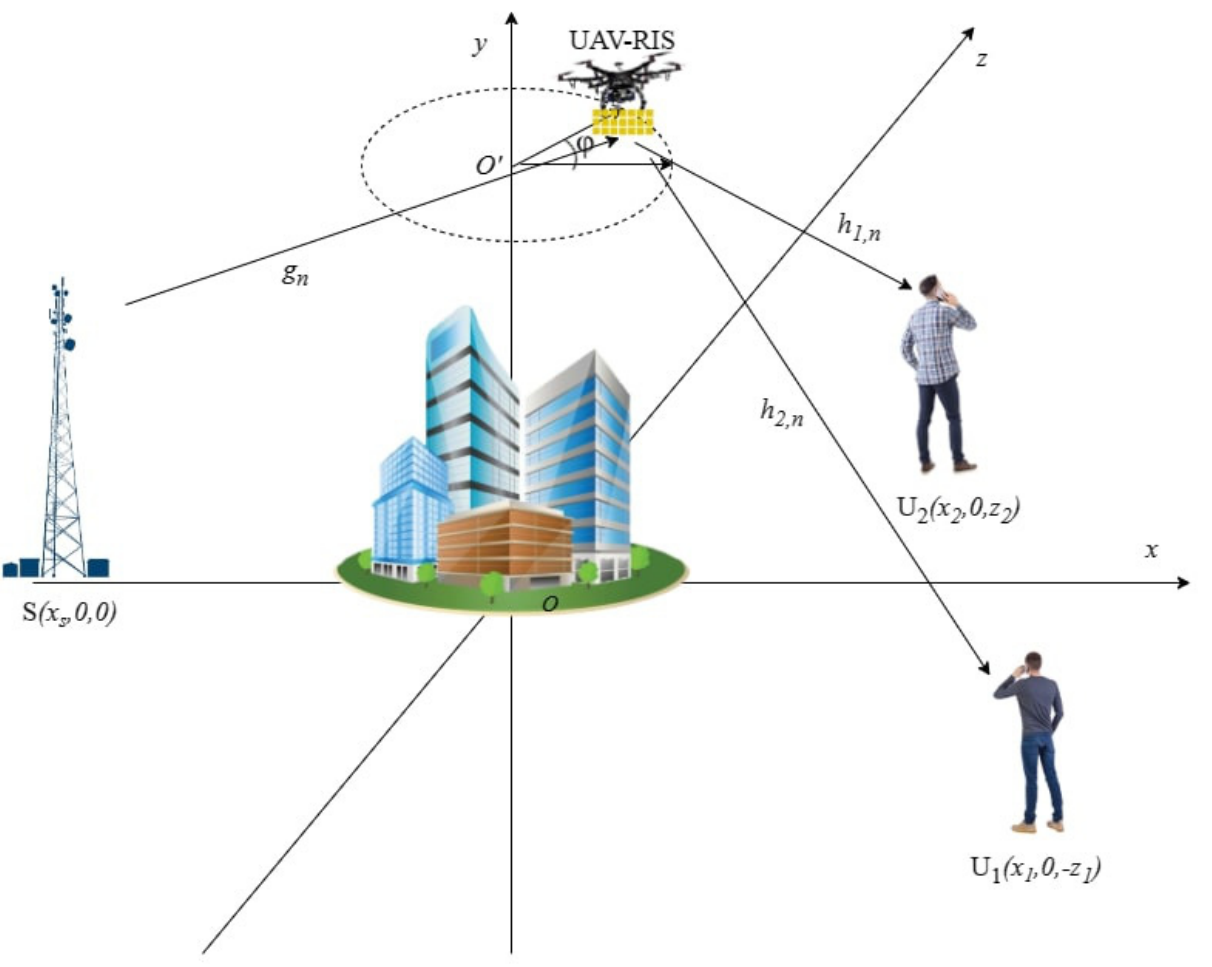}
       \caption{UAV-RIS NOMA system model.}
    \label{constellations}
\end{figure}

Based on NOMA, we assume that the users are ordered based on the {\color{black}end-to-end (e2e)} channel\footnote{{\color{black}The e2e channel refers to the end-to-end channel between S and users, through the UAV-RIS.}} gains as $|q_1|^2<|q_2|^2$, where $q_1$ and $q_2$ can be given as {\color{black}\begin{math} q_i= \sum^N_{n=1} \sqrt{d_{s,n}^{-\alpha} d_{i,n}^{-\alpha}} \ g_n \ \exp{(j\phi_{i,n})} \ h_{i,n} \end{math}}
, where {\color{black} \begin{math} g_n=\nu_n \exp{(-j\theta_n)}\end{math}, \begin{math} h_{i,n}=\mu_{i,n} \exp{(-j\psi_{i,n})}\end{math} are Rayleigh fading channels between S-UAV, and UAV-U$_i$, $i=\{1,2\}$}, $\phi_{i,n}$ represents the adjustable phase induced by the $n^{th}$ reflector of the RIS, {\color{black}$\nu_n$ and $\mu_{i,n}$ are independently Rayleigh distributed random variables,} $d_{s,n}$ {\color{black}and $d_{i,n}$} represent the distances between S and the projection of the $n^{th}$ reflective elements on the ground users, and $\alpha$ is the path-loss exponent{\color{black}, which can be given as \cite{bariah2023performance} $\alpha=\ell_1 Plos(\emptyset)+ \mho_1$, where $Plos(\emptyset)$ is the probability that there will be a LoS component between S-UAV and UAV-users, which is given as \cite{bariah2023performance} $Plos(\emptyset)=[1+ \ell_2 \exp{-\mho_2(\emptyset-\ell_2)}]^{-1}$, where $\ell_1$, $\ell_2$, $\mho_1$ and $\mho_2$ are environment and transmission frequency dependent parameters, and $\emptyset$ is the elevation angle between the S and the RIS reflecting elements and between the RIS reflecting elements and users in radian.} Without loss of accuracy, it is given that $d_{s,n}=d_{s}$ {\color{black}and $d_{i,n}=d_{i}$}. 
Based on the above assumptions, S transmits a superimposed coding signal to the users through the UAV-RIS, which is given as \begin{math}
{\color{black}X=\sqrt{\epsilon_1}s_1 + \sqrt{\epsilon_2} s_2}\end{math}, {\color{black}$\epsilon_1 > \epsilon_2$}, in which ${\color{black}\epsilon_1 + \epsilon_2 }= 1$, {\color{black} where $\epsilon_1$ and $\epsilon_2$ are power allocation factors for U$_1$ and U$_2$, $s_1$ and $s_2$ are the signals\footnote{{\color{black}To accurately analyze the performance of the NOMA scheme with varying power allocations, $s_1$ and $s_2$ are normalized using the unit average power.}} of the U$_1$ and U$_2$}. Considering the HWI and ICI, the received signal reflected through the UAV-RIS at NOMA users can be expressed as
\begin{equation}\label{eq:1}
\begin{split}
y_i= \left( \sqrt{P_{s}}X  + \eta_{t,i} \right)q_i + \eta_{r,i} + I_i + w, \ \ \ {\color{black}i=\{1,2\}}
\end{split}
\end{equation}
{\color{black}where $I_i$ is the ICI of transmissions of neighboring cells. Since we assume that the nodes of neighboring suffer from HWI, $I_i$ can be defined as}
\begin{equation}\label{eq:1}
\begin{split} I_i= \sum_{j \neq i}^{L} \left(\left( \sqrt{P_{j}}X  + \eta_{t,j} \right)q_j + \eta_{r,j} \right), 
\end{split}
\end{equation}
where {\color{black}$L$ represents the number of transmissions of the neighboring cells}, \begin{math}q_j= \sqrt{d_{sj}^{-\alpha} d_{j}^{-\alpha}} \ \sum^N_{n=1}  \ g_{n,j} \ \exp{(j\phi_{n,j})} \ h_{n,j} \end{math}, $d_{sj}$ is the distance between S and UAV-RIS of the $j^{th}$ cells, $d_{j}$ represents the distance between UAV-RIS of the $j^{th}$ cells and U$_i$, \begin{math} g_{n,j}=\nu_{n,j} \exp{(-j\theta_{n,j})}\end{math} is the Rayleigh fading channel between S-UAV of the $j^{th}$ cells, \begin{math} h_{n,j}=\mu_{n,j} \exp{(-j\psi_{n,j})}\end{math} is the Rayleigh fading channel between UAV of the $j^{th}$ cells and $U_i$, $\nu_{n,j}$ and $\mu_{n,j}$ are independently Rayleigh distributed random variables of $j^{th}$ cells, $\phi_{n,j}$ represents the adjustable phase induced by the $n^{th}$ reflector of RIS of the $j^{th}$ cells, $\eta_{t,i} \sim \mathcal{C N} (0, P_{s} {\color{black}k_{t,i}^2)}$ and $\eta_{r,i} \sim \mathcal{C N} (0, P_{s} {\color{black}k_{r,i}^2} |q_i|^2)$ are distortion noises at the transmitter and receiver of the HWI \footnote{{\color{black}In practical applications, transceivers encounter various HWIs. The impact of these impairments can be effectively verified and analyzed by containing an additive independent distortion noise as in \cite{beddiaf2022unified,canbilen2022performance}.}}, respectively, $\eta_{t,j} \sim \mathcal{C N} (0, P_{j} {\color{black}k_{t,j}^2)}$ and $\eta_{r,j} \sim \mathcal{C N} (0, P_{j} {\color{black}k_{r,j}^2} |q_j|^2)$ are distortion noises at the transmitter and receiver of $j^{th}$ cells of the HWI respectively, {\color{black}$k_{t,i}^2$ and $k_{r,i}^2$ are the HWI level at the transceivers, $k_{t,j}^2$ and $k_{r,j}^2$ are the HWI level of $j^{th}$ cells at the transceivers}, $P_{s}$ is the transmit power of S, $P_{j}$ is the total transmit power of $j^{th}$ cells, and $w \sim \mathcal{C N} (0, \sigma^2)$ is the additive white Gaussian noise (AWGN). 

Consequently, to maximize the signal-to-noise ratio (SNR) using \begin{math} \phi_{i,n}= \theta_n+\psi_{i,n} \end{math} and \begin{math} \phi_{n,j}= \theta_{n,j}+\psi_{n,j} \end{math} by adjusting the reflector phases, (1) can be rewritten as 
\begin{equation}\label{eq:1}
\begin{split}
y_i= \sqrt{d_{s}^{-\alpha} d_{i}^{-\alpha}} \ \sqrt{P_{s}} \ \sum^N_{n=1} \ \nu_n \ \mu_{i,n} X + W_i ,
\end{split}
\end{equation}
where \begin{math}W_i=z_1 X k_{t,i} + z_1 k_{r,i} + \sum_{j \neq i}^{L} (X z_2 + z_2 ( k_{t,j}+ k_{r,j} ) )+ w\end{math},
{\color{black}\begin{math}z_1=\sqrt{d_{s}^{-\alpha} d_{i}^{-\alpha}}  \sqrt{P_{s}}  \  \sum^N_{n=1} \nu_n \mu_{i,n}  \end{math}}, \begin{math}z_2=\sqrt{d_{sj}^{-\alpha} d_{j}^{-\alpha}}  \sqrt{P_{j}}  \sum^N_{n=1} \mu_{n,j} \nu_{n,j}\end{math}, and $W_i$ has zero-mean with the variance \begin{math}\sigma_{W,i}^2 \end{math} of U$_i$.

The maximum likelihood detection (MLD) of $s_1$ symbols at the U$_1$ is designed for the RIS scheme by pretending $s_2$’s symbols as noise, and it is given by
\begin{equation}\label{eq:1}
\begin{split}
\Tilde{s}_1= \arg\min_{m} \{|y_1 - \sqrt{d_{s}^{-\alpha} d_{1}^{-\alpha}} \ \sqrt{P_{s}} \ \sum^N_{n=1} \ {\color{black} \nu_{n} \ \mu_{1,n} s_{1,m}|^2}\}  .
\end{split}
\end{equation}

Likewise, the MLD of $s_1$ symbols at the U$_2$ is created for the RIS scheme by portraying $s_2$'s symbols as noise, and it is provided by
\begin{equation}\label{eq:1}
\begin{split}
\Tilde{s}_1= \arg\min_{m} \{|y_2 - \sqrt{d_{s}^{-\alpha} d_{2}^{-\alpha}} \ \sqrt{P_{s}} \ \sum^N_{n=1} \ {\color{black}\nu_{n} \ \mu_{2,n} s_{1,m}|^2}\}  .
\end{split}
\end{equation}

However, in order to detect the $s_2$ at U$_2$, an SIC should be implemented. The MLD of $s_2$ symbols at the U$_2$ for the UAV-RIS NOMA scheme can be given as
\begin{equation}
\begin{split}
\Tilde{s}_2= \arg\min_{m} \{|\acute{y}_2 - \sqrt{d_{s}^{-\alpha} d_{2}^{-\alpha}} \ \sqrt{P_{s}} \ \sum^N_{n=1} \ {\color{black}\nu_{n} \ \mu_{2,n} s_{2,m}|^2}\}  ,
\end{split}
\end{equation}
where 
\begin{equation}
\begin{split}
\acute{y}_2= y_2 - \sqrt{d_{s}^{-\alpha} d_{1}^{-\alpha}} \ \sqrt{P_{s}} \ \sum^N_{n=1} \ {\color{black}\nu_{n} \ \mu_{1,n} {\color{black}\tilde{s}_{1,m}}}  ,
\end{split}
\end{equation}
{\color{black}where $s_{i,m}$ denotes the $m^{th}$ constellation point of U$_i$.} 

\section{Error Probability analysis}
In this section, we analyze the error probability of UAV-RIS NOMA considering both the ideal and non-ideal conditions (i.e., with and without HWI and ICI). The presence of HWI and ICI affect the SIC process and the BER performance. Therefore, the imperfect SIC is considered in the proposed system. In particular, the average BER and then the upper bound are obtained.

Since $\nu_n$, $\mu_{i,n}$, $\nu_{n,j}$, and $\mu_{n,j}$ of (1) and (2) are independently Rayleigh distributed random variables, we put $\mathsf{A_1}=\nu_n \mu_{i,n}$ and $\mathsf{\mathsf{A_2}}=\nu_{n,j}\mu_{n,j}$. Therefore, $\mathbb{E}[\mathsf{A_1}]$=$\mathbb{E}[A_2]$=$\frac{\pi}{4}$ and $\mathsf{VAR[A_1]=VAR[A_2]}=(1-\frac{\pi^2}{16})$. For a sufficiently large number of reflecting elements $N >>1$, according to the central limit theorem, $\mathsf{A_1}$ and $\mathsf{A_2}$ follows Gaussian distribution with parameters, $\mathbb{E}[\mathsf{A_1}]$=$\mathbb{E}[\mathsf{A_2}]$=$N\frac{\pi}{4}$ and $\mathsf{VAR[A_1]=VAR[A_2]}=N(1-\frac{\pi^2}{16})$ {\color{black}\cite{canbilen2022performance}}. Assuming the HWI and ICI, the BER of the UAV-RIS for U$_1$ is obtained by using the decision rule provided in (4) as 
\begin{equation}\label{eq:17}
P_{1} (e)=\frac{1}{2} \sum_{{s}=1}^{2} \mathrm{Q} \left(\sqrt{ \frac{ P_s {\color{black}\zeta_{s}} \psi_1 {\color{black}\lambda}}{\sigma_{W,1}}  } \right),
\end{equation} 
where {\color{black}$\zeta_{s}=[(\sqrt{\epsilon_1}+\sqrt{\epsilon_2}),(\sqrt{\epsilon_1}-\sqrt{\epsilon_2})]$,} $\psi_i=d_{s}^{-\alpha} d_{i}^{-\alpha}$, {\color{black}and $\lambda=|A_1|^2=|\sum^N_{n=1} \nu_n \mu_{i,n}|^2$}.  
The variance $\sigma_{W,1}$ contains the transmitted superimposed coding symbols, therefore there are two different levels of power in binary phase-shift keying (BPSK). Hence, the variance $\sigma_{W,1}$ for U$_1$ can be formulated as {\color{black}
\begin{equation}\label{eq:1}
\begin{split}
& \sigma_{W,1}= \psi_1 \varpi_1 K_a^2   +  \zeta_{s} \aleph  K_b^2 + \aleph K_b^2 + \sigma^2 ,
 \end{split}
\end{equation} }
where, \begin{math}\iota=d_{sj}^{-\alpha} d_{j}^{-\alpha}\end{math}, $K_a^2=k_{t,i}^2+\ k_{r,i}^2$, and $K_b^2=k_{t,j}^2 +\ k_{r,j}^2$ {\color{black} are the aggregate level of HWI, as described in \cite{beddiaf2022unified}}. Based on the central limit theorem, it can be said that $\mathbb{E}[|\sum^N_{n=1} \ \nu_n \ \mu_{i,n} |^2]$ and $\mathbb{E}[|\sum^N_{n=1}\ \nu_{n,j} \ \mu_{n,j}|^2]$ of $\sigma_{W,1}$ in (9) follow a zero-mean Gaussian distribution for $N$ number of reflectors of $j^{th}$ cells. We can obtain $\mathbb{E}[|\sum^N_{n=1} \ {\color{black}\nu_n \ \mu_{i,n}}|^2]=N(1-\frac{\pi^2}{16})$ and $\mathbb{E}[|\sum^N_{n=1}\ \nu_{n,j} \ \mu_{n,j}|^2]=N(1-\frac{\pi^2}{16})$. We use the moment-generating function (MGF) as defined in \cite{canbilen2022performance} of $A_1$, which follows the non-central chi-square distribution, so the average BER for BPSK is given by
{\color{black}\begin{equation}\label{eq:17}
\begin{split}
& P_{1} (e)= \frac{1}{2\pi} \ \sum_{{s}=1}^{2} \int_{0}^{\frac{\pi}{2}} \ \left( \frac{1}{\varpi_{1,a}} \right)^{\frac{1}{2}} \exp{\left(\frac{-\Upsilon_{1,a}}{\varpi_{1,a}}\right)} \text{d} \eta.
\end{split}
\end{equation}
}

Likewise, the analytical BER for U$_2$ is based on the correct and erroneous SIC. Hence, supposing the HWI and ICI, we obtain the average BER of the UAV-RIS for U$_2$ by using the decision rules provided in (6) and (7) as 
\begin{equation}\label{eq:17}
P_{2} (e)=\frac{1}{2} \sum_{{f}=1}^{6} \ \Xi_{f} \ \mathrm{Q} \left(\sqrt{ \frac{ P_s {\color{black}\zeta_{f}} \psi_2 {\color{black}\lambda}}{\sigma_{W,2}}  } \right) ,
\end{equation} 
where $\Xi_{f}=[1,1,-1,1,1,-1]$ and {\color{black}$\zeta_{f}=[\epsilon_2, \epsilon_2, (\sqrt{\epsilon_1}+\sqrt{\epsilon_2}), (\sqrt{2 \epsilon_1}+\sqrt{\epsilon_2}),(\sqrt{\epsilon_1}-\sqrt{\epsilon_2}), (2 \sqrt{\epsilon_1}-\sqrt{\epsilon_2})]$}. The variance $\sigma_{W,2}$ contains the transmitted superimposed coding symbols, thus there are two different levels of power in BPSK. Hence, the variance $\sigma_{W,2}$ for U$_2$ can be formulated as
{\color{black}
\begin{equation}\label{eq:1}
\begin{split}
& \sigma_{W,2}=\psi_2  \varpi_2 K_a^2   + \Delta_{f} \aleph K_b^2+  \aleph K_b^2 + \sigma^2,
 \end{split}
\end{equation}}
 where {\color{black}\begin{math}\varpi_i=P_{s}|\sum^N_{n=1}\nu_n \mu_{i,n}|^2\end{math}, \begin{math}\aleph=\sum_{j \neq i}^{L} \iota  P_{j} |\sum^N_{n=1} \nu_{n,j} \mu_{n,j}|^2\end{math}, and \begin{math}\Delta_{f}=[(\sqrt{\epsilon_1}-\sqrt{\epsilon_2}), (\sqrt{\epsilon_1}+\sqrt{\epsilon_2}), (\sqrt{\epsilon_1}+\sqrt{\epsilon_2}), (\sqrt{ \epsilon_1}+\sqrt{\epsilon_2}),(\sqrt{\epsilon_1}-\sqrt{\epsilon_2}), ( \sqrt{\epsilon_1}-\sqrt{\epsilon_2})]\end{math}.} 
 
 By using the same method for U$_1$, we calculate the average BER for U$_2$ for BPSK as follow
{\color{black}
\begin{equation}\label{eq:17}
\begin{split}
& P_{2} (e)= \frac{1}{2\pi} \ \sum_{{f}=1}^{6} \ \Xi_{f} \ \int_{0}^{\frac{\pi}{2}} \ \left( \frac{1}{ \varpi_{2,a}}\right)^{\frac{1}{2}} \exp{\left(\frac{- \Upsilon_{2,a}}{\varpi_{2,a}}\right)} \text{d} \eta,
\end{split}
\end{equation}
where $\varpi_{i,a}=1+\mho_{i,a} \Lambda$, $\Upsilon_{i,a}=\wp_{i,a} \Lambda$, $\mho_{i,a}=\frac{N(16-\pi^2) P_s \psi_i}{8 \sin^2(\eta)\sigma_{W,i}}$,
$\wp_{i,a}=\frac{N^2 \pi^2 P_s \psi_i}{16 \sin^2(\eta)\sigma_{W,i}}$, and $\Lambda$ is equal to $\zeta_{s}$ for BER of U$_1$ and equal to $\zeta_{f}$ for BER of U$_2$.
}

To acquire insights, we compute the upper bound of (10) and (13), respectively, by setting $\eta=\pi/2$ as {\color{black}
\begin{equation}\label{eq:17}
\begin{split}
& P_{i} (e) \leq \frac{1}{4} \ \sum_{{r}=1}^{\vartheta} \ \partial \ \left( \frac{1}{\varpi_{i,b}} \right)^{\frac{1}{2}} \exp{\left(\frac{-\Upsilon_{i,b}}{\varpi_{i,b}}\right)} , 
\end{split}
\end{equation} 
where $\varpi_{i,b}=1+\mho_{i,b} \Lambda$, $\Upsilon_{i,b}=\wp_{i,b} \Lambda$, $\mho_{i,b}=\frac{N(16-\pi^2) P_s \psi_i}{8 \sigma_{W,i}}$,
$\wp_{i,b}=\frac{N^2 \pi^2 P_s \psi_i}{16 \sigma_{W,i}}$, $\vartheta$ is equal to 2 for BER of U$_1$ and equal to 6 for BER of U$_2$, and $\partial$ is equal to 1 for U$_1$ and equal to $\Xi_{f}$ for U$_2$. 
}
{\color{black}This result leads to the following observation}. The error performance of UAV-RIS NOMA improves with increasing $N$. However, it is limited by the imperfection of the non-ideal conditions (i.e., HWI and ICI) that cause an error floor at high SNR regions.


In our considered system, we provide the BER of system, which is the error probability when the U$_1$, U$_2$, or both detect their own symbols incorrectly. Hence, the BER of system for UAV-RIS NOMA with HWI and ICI can be expressed as \cite{beddiaf2022unified}
\begin{equation}\label{eq:17}
\begin{split}
& P_{system} (e)= 1- (1-P_{1} (e)) (1-P_{2} (e)),
\end{split}
\end{equation}
where $P_{1} (e)$ and $P_{2} (e)$ are the BER of U$_1$ and U$_2$, respectively.

 \section{Numerical Results} 
In this section, we present the Monte-Carlo simulations to validate the derivations of the previous sections. We discuss the effect of HWI, ICI, and imperfect SIC on the error performance of UAV-RIS NOMA. Without loss of generality, we consider that $k=K_a^2=K_b^2$, $P=P_{s}=P_{j}$, and the total transmit SNR is defined as $\gamma=\frac{P}{\sigma^2}$. The parameters used in all simulations
are given by{\color{black}\cite{bariah2023performance}} $r=5$ m, $T=20$ m, $\varphi=\frac{\pi}{4}$, $\mathsf{S}(-5,0,0)$, $\mathsf{U_1}(10,0,-10)$, $\mathsf{U_2}(5,0,5)$, $d_{sj}=100$ m, $d_{j}=100$ m, {\color{black}$\ell_1=-1.5$, $\mho_1=3.5$, $\ell_2 =2$, $\mho_2=0.16$, $\emptyset=1.2$,} $k=0, 0.15$, $L=0, 3$, the computer simulation repeated $10^5$ iterations, and $\epsilon_2=0.1, 0.2$. 

\begin{figure}
    \centering
    \subfloat[{}]{\includegraphics[width=1\columnwidth]{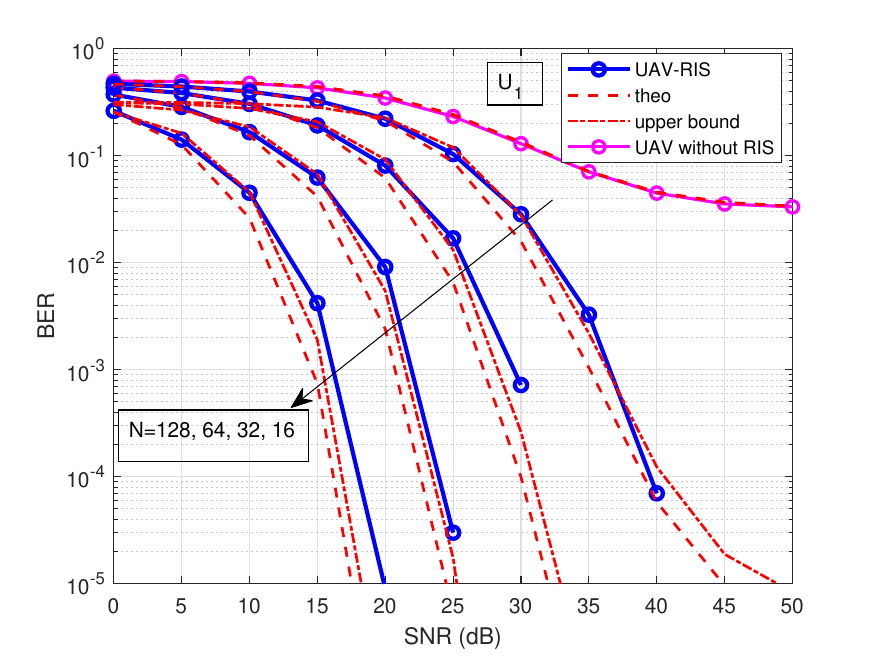}}
  \\
 \centering
    \subfloat[{}]{\includegraphics[width=1\columnwidth]{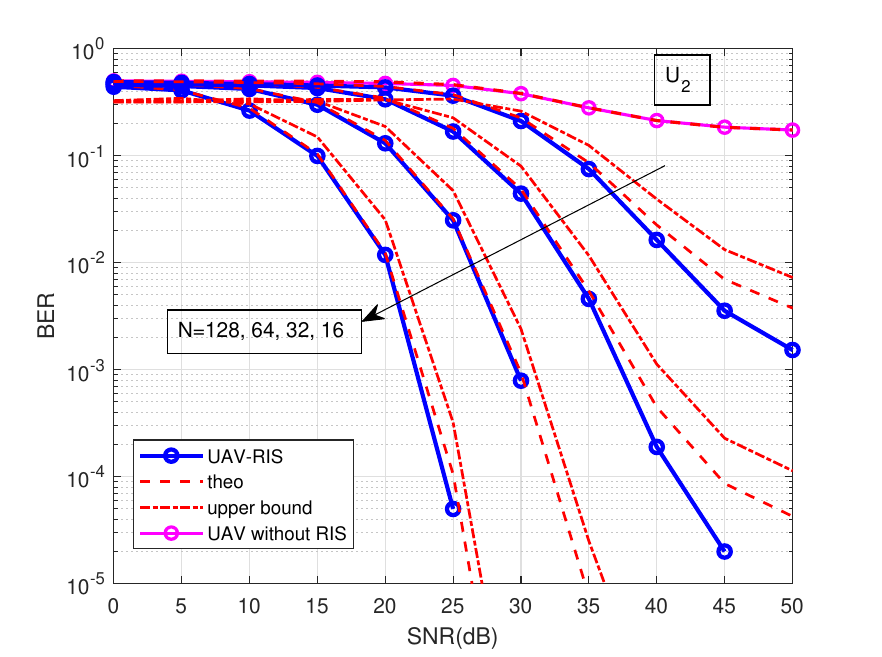}}
    \caption{BER w.r.t. SNR of UAV-RIS NOMA with the fixed value of the effect of HWI and ICI: (a) U$_1$; (b) U$_2$.}
    \label{constellations}
\end{figure}

\begin{figure}
    \centering
    \subfloat[{}]{\includegraphics[width=1\columnwidth]{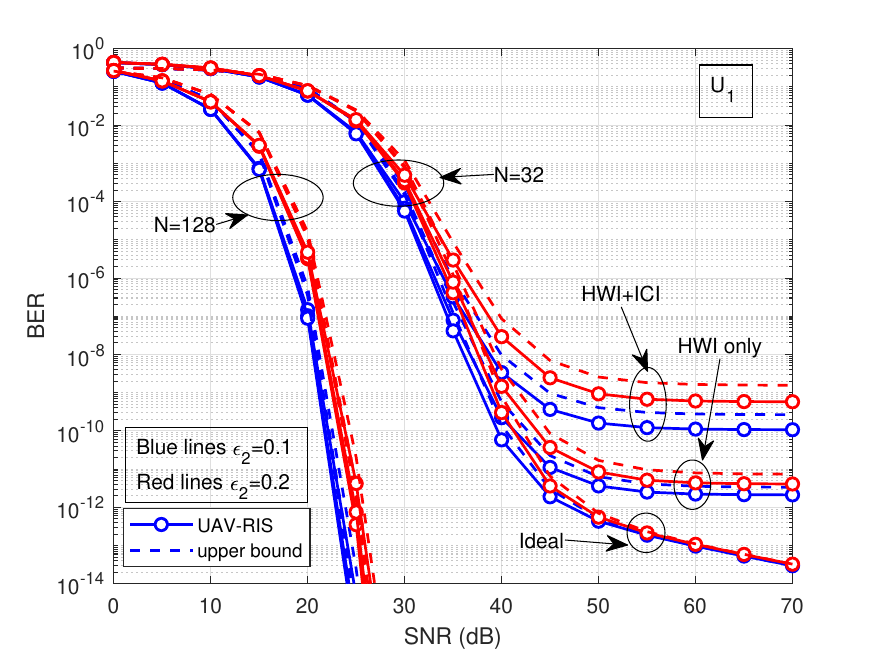}}\\
 \centering
    \subfloat[{}]{\includegraphics[width=1\columnwidth]{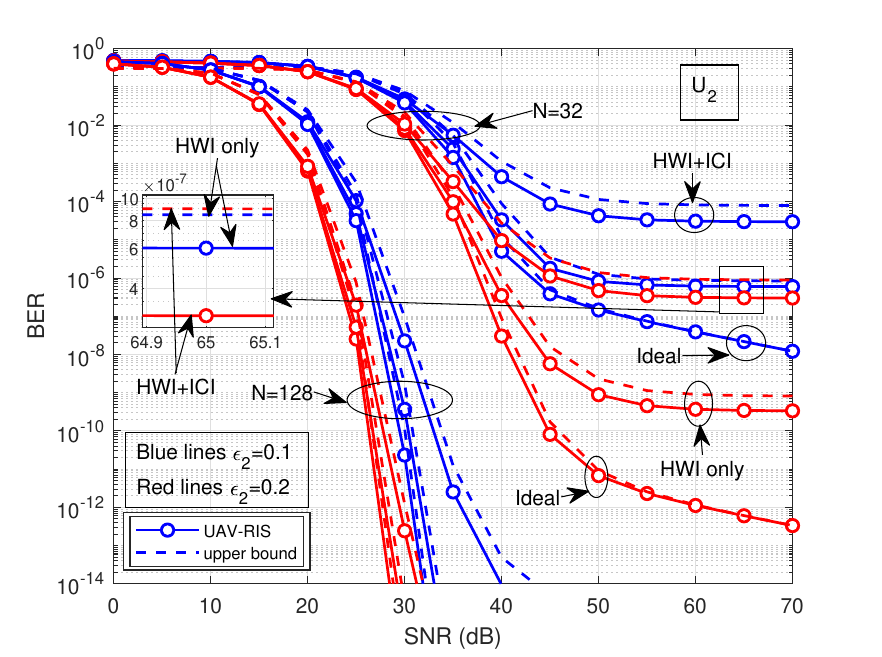}}
    \caption{BER w.r.t. SNR of UAV-RIS NOMA with the effect of HWI and ICI: (a) U$_1$; (b) U$_2$.}
    \label{constellations}
\end{figure}



In Fig. 2, we present the BER performance of UAV-RIS for two NOMA users with the impact of HWI and ICI with fixed $k=0.15$ and $L=3$. We observe that the analytical results perfectly matched with simulation results. The upper bound is quite strict because the analytical and simulation results match for $N\geq64$. It can be also seen that increasing the number of elements $N$ provides a better BER performance. Clearly, the performance of UAV-RIS is superior to that of UAV without RIS scheme with increasing $N$ and achieves a performance gain of more than 5 dB with less error at lower SNRs. This figure is extended to Fig. 3 for more investigation into the impact of ideal and non-ideal conditions (i.e., with{\color{black}/}without HWI and ICI) using the analytical expressions. 
It is observed from this figure that the BER curves have a waterfall region and a saturation region in the ideal and non-ideal cases at high SNR values. This observation can be explained as follows. According to {\color{black} (14)}, for the low values of $N \frac{P}{\sigma_{W,i}}$, we achieve an incredibly low BER performance by increasing $N$ due to $N^2$ in the exponential form. On the other side, for the high values of $N \frac{P}{\sigma_{W,i}}$, the BER performance is saturated due to the $\frac{-1}{2}$ exponent in the high SNR values. Moreover, the presence of HWI and ICI have much impact on the system performance in the high SNR values, especially for U$_2$ more than U$_1$ due to the SIC process. It is observed that the BER is nearly the same for both ideal and non-ideal conditions until a particular SNR value. Therefore, there is a trade-off between the number of $N$ and the level of SNR to enable the impacts of the HWI and ICI to be reasonably disregarded. Despite the significant impact of the HWI and ICI which leads to creating an error floor at the higher SNR values, the reached BER at higher values of $N$ is gratifying and sufficient to ignore the influence of HWI and ICI at low SNR values {\color{black} to maintain high energy efficiency}. Moreover, the change in the power allocation does not improve the performance while it affects one user at the expense of the other. It affects the SIC and the detection of the signals.





\begin{figure}
\centering
\includegraphics[width=.95\columnwidth]{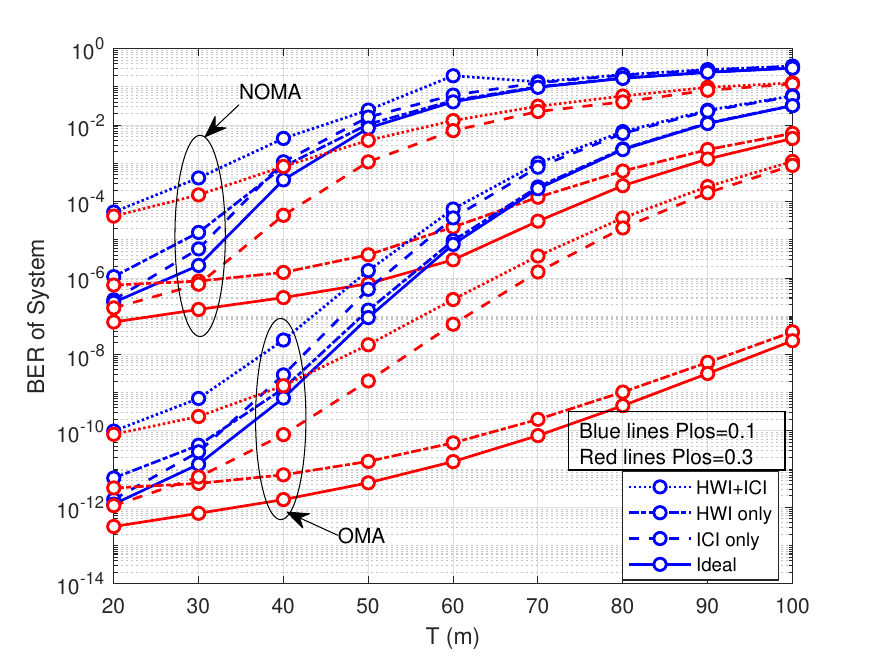}
       \caption{BER of the system from (15) w.r.t. altitude (T (m)) of UAV-RIS NOMA with the effect of HWI and ICI.}
    \label{constellations}
\end{figure}

In order to show the role of UAV, in Fig. 4, we present the BER of system {\color{black}from (15) versus $T (m)$ with different LoS probability for }UAV-RIS NOMA compared to OMA with the effect of HWI and ICI when SNR=50 dB, $N$=32{\color{black}, and $Plos(\emptyset)=0.1, 0.3$}. In this figure, it is assumed that the time-division multiple access (TDMA) is implemented for OMA, where each user transmits its signals at different times. It is observed that increasing $T$ values degrades the BER performance of the system. It is due to increasing the Euclidean distance between the S and UAV and between UAV and users. In the lower values of $T$, when the HWI and ICI have a joint effect, there is more imperfection in the system compared to HWI only and ICI only. {\color{black}The impact of ICI exceeds that of HWI at higher $T$ values. It can be explained as follows. Increasing the altitude of a UAV may increase its proximity to neighboring UAVs, which means increasing the effect of ICI.} In higher values of $T$, the impact of HWI and ICI are relatively matched to the ideal case due to the increasing error to detect the symbols in the higher Euclidean distances. {\color{black}It can also be observed that increasing the LoS component value also means improving direct visibility between the transmitter and the receiver, which reduces path loss and enhances performance. }It can be also seen that the OMA has better performance gain compared to NOMA. Combining the users' signals in NOMA makes it suffer from inter-user interference (IUI), which complicates the SIC performance to detect the users' signals. Moreover, the presence of HWI and ICI impedes the SIC process and causes more errors in the detection of the users' signals. On the other side, OMA transmits the signals at different times, therefore it does not suffer from IUI.

\section{Conclusion}
In this letter, the error probability is analyzed for UAV-RIS NOMA. We consider that practical constraints such as HWI, ICI, and imperfect SIC exist in our proposed system. We discussed the effects of these imperfections on the system's performance. The upper bound of the analytical derivation is obtained and the analytical derivations are validated by Monte Carlo simulations. The results show that the non-ideal conditions (HWI, ICI, and imperfect SIC) cause a degradation in the system performance at high SNR regions. However, these degradations can be ignored by increasing the number of reflecting elements at low SNR because it achieves satisfying results{\color{black}, which leads to the preservation of high energy efficiency}. Moreover, the proposed system has a higher performance gain compared to the UAV without RIS, while OMA has better performance than NOMA because it does not suffer from IUI. This letter provides a comprehensive examination of the practical constraints for UAV-RIS NOMA systems, giving researchers a clearer vision of these flaws conditions for future research. This letter, however, opens up future research on multiple-input multiple-output (MIMO) for UAV-RIS.




\bibliographystyle{IEEEtran}
\bibliography{references}

\begin{thebibliography}{10}
\providecommand{\url}[1]{#1}
\csname url@samestyle\endcsname
\providecommand{\newblock}{\relax}
\providecommand{\bibinfo}[2]{#2}
\providecommand{\BIBentrySTDinterwordspacing}{\spaceskip=0pt\relax}
\providecommand{\BIBentryALTinterwordstretchfactor}{4}
\providecommand{\BIBentryALTinterwordspacing}{\spaceskip=\fontdimen2\font plus
\BIBentryALTinterwordstretchfactor\fontdimen3\font minus \fontdimen4\font\relax}
\providecommand{\BIBforeignlanguage}[2]{{%
\expandafter\ifx\csname l@#1\endcsname\relax
\typeout{** WARNING: IEEEtran.bst: No hyphenation pattern has been}%
\typeout{** loaded for the language `#1'. Using the pattern for}%
\typeout{** the default language instead.}%
\else
\language=\csname l@#1\endcsname
\fi
#2}}
\providecommand{\BIBdecl}{\relax}
\BIBdecl

\bibitem{alfattani2021aerial}
S.~Alfattani, W.~Jaafar, Y.~Hmamouche, H.~Yanikomeroglu, A.~Yonga{\c{c}}oglu, N.~D. D{\`a}o, and P.~Zhu, ``{Aerial platforms with reconfigurable smart surfaces for 5G and beyond},'' \emph{IEEE Commun. Mag.}, vol.~59, no.~1, pp. 96--102, 2021.

\bibitem{bariah2023performance}
L.~Bariah, F.~Boukhalfa, W.~Jaafar, S.~Muhaidat, and H.~Yanikomeroglu, ``{On the performance of RIS-enabled NOMA for aerial networks},'' in \emph{2023 IEEE Wirel. Commun. Net. Conf. (WCNC)}, pp. 1--6.

\bibitem{agrawal2021performance}
N.~Agrawal, A.~Bansal, K.~Singh, and C.-P. Li, ``{Performance evaluation of RIS-assisted UAV-enabled vehicular communication system with multiple non-identical interferers},'' \emph{IEEE Trans. Intell. Transp. Sys.}, vol.~23, no.~7, pp. 9883--9894, 2021.

\bibitem{solanki2022performance}
S.~Solanki, J.~Park, and I.~Lee, ``{On the performance of IRS-aided UAV networks with NOMA},'' \emph{IEEE Trans. Veh. Tech.}, vol.~71, no.~8, pp. 9038--9043, 2022.

\bibitem{li2021aerial}
Y.~Li, C.~Yin, T.~Do-Duy, A.~Masaracchia, and T.~Q. Duong, ``{Aerial reconfigurable intelligent surface-enabled URLLC UAV systems},'' \emph{IEEE Access}, vol.~9, pp. 140\,248--140\,257, 2021.

\bibitem{nguyen2022uav}
M.-H.~T. Nguyen, E.~Garcia-Palacios, T.~Do-Duy, O.~A. Dobre, and T.~Q. Duong, ``{UAV-aided aerial reconfigurable intelligent surface communications with massive MIMO system},'' \emph{IEEE Trans. Cog. Commun. Net.}, vol.~8, no.~4, pp. 1828--1838, 2022.

\bibitem{canbilen2022performance}
A.~E. Canbilen, E.~Basar, and S.~S. Ikki, ``{On the performance of RIS-assisted space shift keying: Ideal and non-ideal transceivers},'' \emph{IEEE Trans. Commun.}, vol.~70, no.~9, pp. 5799--5810, 2022.

\bibitem{beddiaf2022unified}
S.~Beddiaf, A.~Khelil, F.~Khennoufa, F.~Kara, H.~Kaya, X.~Li, K.~Rabie, and H.~Yanikomeroglu, ``{A unified performance analysis of cooperative NOMA with practical constraints: Hardware impairment, imperfect SIC and CSI},'' \emph{IEEE Access}, vol.~10, pp. 132\,931--132\,948, 2022.

\bibitem{zhou2023outage}
M.~Zhou, Y.~Li, Y.~Sun, and Z.~Ding, ``{Outage performance of RIS-assisted V2I communications with inter-cell interference},'' \emph{IEEE Wirel. Commun. Lett.}, vol.~12, no.~6, pp. 962--966, 2023.

\bibitem{singh2022noma}
S.~K. Singh, K.~Agrawal, K.~Singh, C.-P. Li, and M.-S. Alouini, ``{NOMA enhanced UAV-assisted communication system with nonlinear energy harvesting},'' \emph{IEEE Open J. Commun. Soc.}, vol.~3, pp. 936--957, 2022.

\end{thebibliography}

\vfill

\end{document}